\documentclass[12pt]{article}
\usepackage{amsthm}
\usepackage{eurosym}
\usepackage{amsfonts}
\usepackage{amssymb,amsmath}
\usepackage[dvips]{color}
\usepackage{amscd}
\usepackage{tikz}
\usepackage{mathtools}
\usepackage{graphicx}
\usepackage{caption}
\usepackage{subcaption}
\usepackage[all,cmtip]{xy}

\usetikzlibrary{decorations.pathreplacing}

\usetikzlibrary{patterns}

\def\and{\mathrm{and}}

\newcommand{\be}{\begin{equation}}
\newcommand{\ee}{\end{equation}}
\newcommand{\bea}{\begin{eqnarray}}
\newcommand{\eea}{\end{eqnarray}}
\newcommand{\beas}{\begin{eqnarray*}}
\newcommand{\eeas}{\end{eqnarray*}}
\newcommand{\ba}{\begin{array}}
\newcommand{\ea}{\end{array}}

\newcommand{\nbox}{{\,\lower0.9pt\vbox{\hrule \hbox{\vrule height 0.2 cm \hskip 0.19 cm \vrule height 0.2 cm}\hrule}\,}}

\def\href#1#2{#2}
\begin{document}

\begin{titlepage}
\hfill
\vbox{
    \halign{#\hfil         \cr
           }
      }

\hbox to \hsize{{}\hss \vtop{ \hbox{}

}}

\vspace*{20mm}

\begin{center}

{\normalsize \textbf{Holographic derivation of a class of short range correlation functions}}

{\normalsize \vspace{10mm} }

{\normalsize {Hai Lin${}^{1}$}, Haoxin Wang${}^{1,2}$ }

{\normalsize \vspace{10mm} }

{\small \emph{${}^1$\textit{Yau Mathematical Sciences Center, Tsinghua University,
Beijing 100084, P. R. China
}} }

{\normalsize \vspace{0.2cm} }

{\small \emph{$^2$\textit{Department of Mathematical Sciences, Tsinghua University,
Beijing 100084, P. R. China
\\
}} }

{\normalsize \vspace{0.4cm} }

\end{center}

\begin{abstract}
We construct a class of backgrounds with a warp factor and anti-de Sitter asymptotics, which are dual to boundary systems that have a ground state with a short-range two-point correlation function. The solutions of probe scalar fields on these backgrounds are obtained by means of confluent hypergeometric functions. The explicit analytical expressions of a class of short-range correlation functions on the boundary and the correlation lengths $\xi$ are derived from gravity computation. The two-point function calculated from gravity side is explicitly shown to exponentially decay with respect to separation in the infrared. Such feature inevitably appears in confining gauge theories and certain strongly correlated condensed matter systems.

\end{abstract}

\end{titlepage}

\vskip 1cm

\section{Introduction}
\label{sec_Introduction}\renewcommand{\theequation}{1.\arabic{equation}} %
\setcounter{equation}{0}

The gauge/gravity correspondence \cite%
{Maldacena:1997re,Gubser:1998bc,Witten:1998qj} has given an extraordinary
method to study a quantum system by a higher dimensional gravity, which
relates a theory with gravity to a quantum system without gravity in a
non-trivial way. This duality indicates the emergence of bulk spacetime
geometry from the degrees of freedom living in the boundary \cite%
{Rangamani:2016dms,VanRaamsdonk:2010pw,Horowitz:2006ct,Koch:2009gq}. It
further provides us a way to compute interesting quantitative features of
strongly-coupled quantum systems and non-perturbative effects of quantum
field theories, since it allows us to make predictions of observables
pertaining to the boundary system, by working in the gravity side.

The gauge/gravity correspondence enables us to compute correlation functions
of a boundary conformal field theory by working in the gravity, and the
details of this procedure were reviewed in \cite{DHoker:2002nbb}. Most of
the gravity computations in the literature are for long-range correlation
functions in the conformal field theory. On the other hand, short-range
correlation functions are also very interesting and important in both
quantum field theories and condensed matter systems. We focus on a class of
short-range correlation functions in a $D$-dimensional system and derive
them from gravity computation, using a new class of gravity backgrounds that
we construct. The appearance of a holographic direction for a boundary
system is\ also closely related to the renormalization group flow of the
boundary system, e. g. \cite{Skenderis:2002wp,Papadimitriou:2004ap}. Our
gravity ansatz is similar to the ansatz used in the aforementioned
holographic renormalization group.

We want to construct a class of new backgrounds that enable us to compute
short-range exponential decay two-point functions around the ground state,
with a correlation length. The short-range correlation is distinguished from
the long-range correlation. The short-range correlation means that the
correlation length is much smaller than the overall size$~$of the boundary
system. If the overall size of the boundary system is infinite, a finite
correlation length will lead to a short-range correlation.

These short-range correlation functions are similar to those that occur in
the vacua of massive quantum field theories. On the other hand, vacua of
massive quantum field theories have gravity dual descriptions, e.g. \cite%
{Polchinski:2000uf,Pope:2003jp,Bena:2004jw,Lin:2004nb,Lin:2005nh,Conde:2016hbg}%
. We compute short-range correlated two-point correlation functions from the
gravity side with the backgrounds in our paper, and our results are also
potentially related to confining gauge theories.

The gauge/gravity correspondence is very useful for studying condensed
matter systems, since it can give descriptions in strong coupling regimes,
e.g. \cite{McGreevy:2009xe,Hartnoll:2016apf,Rangamani:2016dms}. The feature
that the boundary systems in our case have a short-range correlation
function near the ground state, is also similar to that of many strongly
correlated condensed matter systems. These aspects are also very interesting
for investigations.

The organization of this paper is as follows. In Section 2, we construct a
class of backgrounds with a warp factor and anti-de Sitter asymptotics,
which we will show that the dual boundary systems have a short-range
two-point correlation function around the ground state. In Section 3, we
solve basis solutions of a probe scalar field on these curved backgrounds,
by means of confluent hypergeometric functions. Then in Section 4, we
compute the bulk-to-bulk propagator and the boundary-to-bulk propagator in
these backgrounds. Afterward in Section 5, we derive from the gravity side
the short-range correlation function of the boundary system and the
corresponding correlation length. In Section 6, we analyze the implication
of our gravity computation to the short-range correlation and the
correlation length. Finally, we discuss our results and draw some
conclusions in Section 7. In Appendix A, we include details of our
derivation of the background solutions in matter coupled gravity. In
Appendix B, we describe relations between our ansatz and that used in the
holographic renormalization. In Appendices C and D, we include detailed
derivations for the basis solutions and the propagators, respectively.

\section{A class of geometries for short range correlation}
\label{sec_intro_ copy(1)} \renewcommand{\theequation}{2.\arabic{equation}} %
\setcounter{equation}{0} \renewcommand{\thethm}{2.\arabic{thm}} %
\setcounter{thm}{0} \renewcommand{\theprop}{2.\arabic{prop}} %
\setcounter{prop}{0}

We desire a short-range correlation function evaluated around the ground
state with a correlation length $\xi $ for a $D$-dimensional spacetime. This
$D$-dimensional spacetime can be constructed as an asymptotic boundary of a
gravity system in higher dimensions. Let us consider that this gravity
system is on the spacetime that we denote as $M$. The gravity system
contains a holographic direction which we denote by $z$ here. Using a warp
factor $a^{2}(z)$, the metric ansatz of $M$ can have the form
\begin{equation}
ds^{2}=a^{2}(z)(\eta _{\mu \nu }\mathrm{d}x^{\mu }\mathrm{d}x^{\nu }+\mathrm{%
d}z^{2}),  \label{metric_}
\end{equation}%
where $\mu =0,\cdots ,D-1$, and $z>0$ is the holographic radial direction
and $x$ is denoted as the spacetime position vector on the $D$-dimensional
boundary $\partial M$. This metric ansatz (\ref{metric_}) is used
extensively in the holographic analysis of the renormalization group flow of
boundary systems, for a review see e.g. \cite{DHoker:2002nbb}.

We want to construct a class of new backgrounds of the form (\ref{metric_}),
that are dual to a quantum theory on the boundary with exponential decay
two-point function around the ground state, with leading behavior
\begin{equation}
\langle \mathcal{O}(x)\mathcal{O}(x^{\prime })\rangle \sim e^{-|x-x^{\prime
}|/\xi }  \label{correlation_01}
\end{equation}%
when the separation $|x-x^{\prime }|~$is much larger than the correlation
length $\xi $. The short-range correlation is distinguished from the
long-range correlation. Consider\ that the boundary system has an overall
size$~$of the system $l_{\text{sys}}$. The short-range correlation means
that $\xi \ll $ $l_{\text{sys}}$. Hence if the overall size of the boundary
system is infinite, a finite correlation length $\xi $ will lead to a
short-range correlation. In a many-body condensed matter system, $l_{\text{%
sys}}$ is of order the finite size of the material. In this paper, we focus
on a class of short-range correlations and derive them from gravity
computation.

Since the energy scale of the boundary system is related to the inverse of
the radial direction $z$, there must exists a special radius scale $z=z_{0}$
in the gravity dual characterizing the energy scale $\xi ^{-1}$ in the
infrared of the boundary system. On general grounds, we expect that $\xi $
is a function $\xi (z_{0})~$of $z_{0}$, and $\xi $ may also depend on other
parameters.

Here we still work on the asymptotically AdS background, where the
asymptotic boundary is at $z=0$. We consider that $a^{2}(z)~$can be expanded
in powers of $z/L$, namely,
\begin{equation}
a^{2}(z)=\frac{L^{2}}{z^{2}}+\frac{\eta (z_{0})L}{z}+\gamma (z_{0})+O(\frac{z%
}{L}),  \label{metric_01}
\end{equation}%
where $\eta (z_{0})$ and $\gamma (z_{0})$ are functions of $z_{0}$,
abbreviated as $\eta $ and $\gamma $ in this paper. In the later sections,
we will show that above warp factor is a nice choice we desire.

The above metric with the warp factors (\ref{metric_01}) can be solved in
matter coupled gravity. The details of our derivation are in Appendix A. As
an example, they can be obtained in scaler coupled gravity with the action
\begin{equation}
S=\frac{1}{2\kappa ^{2}}\int \mathrm{d}^{D+1}y\sqrt{-g}\left[ R-{\frac{1}{2}}%
\partial _{M}\varphi \partial ^{M}\varphi -V(\varphi )\right] ,
\end{equation}%
where $\kappa $ is the gravitational coupling constant, and $M=0,\cdots ,D.$
The profile of the scalar field $\varphi $ deforms the AdS background.

For%
\begin{equation}
a^{2}(z)=\frac{L^{2}}{z^{2}}+\frac{\eta L}{z}+\gamma +O(\frac{z}{L}),
\label{solution_01_1}
\end{equation}%
to the first three orders in $z$, the scalar and its potential are
\begin{equation}
V=-\frac{D(D-1)}{L^{2}}+(D-1)(2D-1)\frac{\eta z}{L^{3}}+(D-1)(12D\gamma
-12\gamma -13D\eta ^{2}+11\eta ^{2})\frac{z^{2}}{4L^{2}},
\label{solution_01_2}
\end{equation}%
\begin{equation}
\varphi =\varphi _{0}\mp \frac{1}{6}\sqrt{\frac{(D-1)z}{-2\eta L}}\left[
24\eta +(12\gamma -7\eta ^{2})\frac{z}{L}\right] .  \label{solution_01_3}
\end{equation}%
Here we require $\eta <0$. The meaning of $\varphi _{0}$ is that it is the
value of $\varphi $ at $z=0$. Note that, Eq. (\ref{solution_01_2})--(\ref%
{solution_01_3}) gives a parametric form of $V(\varphi )$, where $V~$is a
function of $\varphi $, written in a parametric representation.

We also find an exact solution, with the warp factor
\begin{equation}
a(z)=L\left( \frac{1}{z}-\frac{1}{z+2z_{0}}\right) ,  \label{metric_02}
\end{equation}%
with $z\geq 0$ and $2z_{0}>0$, and the corresponding scalar and its
potential are
\begin{equation}
V(\varphi )=-\frac{(D-1)}{8L^{2}}\left( (2D-1)e^{\frac{\varphi -\varphi _{0}%
}{\sqrt{D-1}}}+(2D-1)e^{\frac{\varphi _{0}-\varphi }{\sqrt{D-1}}%
}+2(2D+1)\right) ,  \label{solution_02_2}
\end{equation}%
\begin{equation}
\varphi =\varphi _{0}\pm 4\sqrt{D-1}\log \left( \sqrt{\frac{z}{2z_{0}}+1}+%
\sqrt{\frac{z}{2z_{0}}}\right) .  \label{solution_02_3}
\end{equation}%
The solution (\ref{metric_02})--(\ref{solution_02_3}) is an exact solution.
If expanded, it is a special case of the solution (\ref{solution_01_1})--(%
\ref{solution_01_3}) for $\eta (z_{0})=-Lz_{0}^{-1}$,$~\gamma (z_{0})=\frac{3%
}{4}L^{2}z_{0}^{-2}$.

In the above analysis, the case for maximally symmetric AdS geometry is\ $V=-%
\frac{D(D-1)}{L^{2}}\mathrm{\ }$and $\varphi =\varphi _{0}$, corresponding
to $\eta =0$ and $\gamma =0.$

Our metrics may be relevant for holographic normalization schemes, e.g. \cite%
{Skenderis:2002wp,Papadimitriou:2004ap} and the relations between our ansatz
and that used in the holographic renormalization are described in Appendix B.

\section{The basis solutions}
\renewcommand{\theequation}{3.\arabic{equation}} \setcounter{equation}{0}

In this section, we consider a probe scalar field $\phi ~$on these curved
backgrounds (\ref{metric_01}), (\ref{metric_02}) whose boundary value is
regarded as the source coupling to $\mathcal{O}(x)$ which has a short-range
correlation function as (\ref{correlation_01}) in the large separation $%
|x-x^{\prime }|$. We do not consider the back-reaction of the probe scalar
field $\phi ~$to the backgrounds. The action of $\phi $ in the curved
background (\ref{metric_01}) reads
\begin{equation}
S\,=\,-{\frac{1}{2}}\,\,\int \mathrm{d}^{D+1}y\sqrt{|g|}[\,g^{MN}\,\partial
_{M}\phi \,\partial _{N}\phi \,+\,m^{2}\phi ^{2}\,]\,,
\label{scalar_action_AdS_01}
\end{equation}%
where $y$ parametrizes the coordinates $\left( z,x^{\mu }\right) $ in $D+1$
dimensions. The equation of motion derived from the action (\ref%
{scalar_action_AdS_01}) is the Klein-Gordon equation, in the form
\begin{equation}
{\frac{1}{\sqrt{|g|}}}\partial _{M}(\sqrt{|g|}g^{MN}\partial _{N}\,\phi
)-\,m^{2}\phi \,=\,0\,.
\end{equation}%
More explicitly, after substituting the metric (\ref{metric_}), it becomes
\begin{equation}
a^{-2}\left( -\partial _{z}^{2}-(D-1)\left( \ln a\right) ^{\prime }\partial
_{z}-\,\partial _{\mu }\partial ^{\mu }+m^{2}a^{2}\right) \phi (z,x^{\mu
})=0.
\end{equation}%
One may perform the Fourier transform of $\phi $ in the $x^{\mu }$
coordinates
\begin{equation}
\phi (z,x^{\mu })\,=\,\int {\frac{\mathrm{d}^{D}k}{(2\pi )^{D}}}%
\,\,e^{ik\cdot x}\,\phi (z,k^{\mu })\,,
\end{equation}%
so that the equation could be rewritten in the form of the Helmholtz
equation
\begin{equation}
\left( -\partial _{z}^{2}-(D-1)\left( \ln a\right) ^{\prime }\partial
_{z}+k^{2}+m^{2}a^{2}\right) \phi (z,k^{\mu })=0,  \label{Helmholtz}
\end{equation}%
where $k^{2}=k_{\mu }k^{\mu }$ and we use the metric (\ref{metric_01}). It
gives the following form
\begin{eqnarray}
&&\left( \left( k^{2}+m^{2}\gamma (z_{0})\right) +\frac{m^{2}L^{2}}{z^{2}}%
+m^{2}\eta (z_{0})\frac{L}{z}\right.   \notag \\
&&\left. -(D-1)\left( -\frac{1}{z}+\frac{\eta (z_{0})}{2L}+\frac{z}{L^{2}}%
\left( \gamma (z_{0})-\frac{\eta (z_{0})^{2}}{2}\right) \right) \partial
_{z}-\partial _{z}^{2}\right) \phi (z,k)=0.  \label{new_Helmholtz_02}
\end{eqnarray}%
This equation has exact solutions if the linear term in the first-order
derivative term is neglected. The two linearly independent basis solutions
are by means of the confluent hypergeometric functions of the second kind $%
U(a,b,x)$ and of the first kind $_{1}F_{1}(a,b,x)$, respectively \cite%
{Abramowitz Stegun,Gradshteyn Ryzhik}. The detailed derivations are in
Appendix C.

Consider this special solution,
\begin{equation}
\phi (z,k)=z^{\frac{D}{2}+\nu }e^{-\frac{(\beta +(D-1)\eta )z}{4L}}U\left(
\alpha +\nu ,2\nu +1,\frac{\beta z}{2L}\right) .  \label{U_solution_02}
\end{equation}%
Here $\nu =\sqrt{\frac{D^{2}}{4}+m^{2}L^{2}}$, and
\begin{equation}
\alpha =\frac{1}{2}-\frac{\eta }{2\beta }\left[ (D-1)^{2}-4m^{2}L^{2}\right]
,
\end{equation}
\begin{equation}
\beta =4L(k^{2}+\xi ^{-2})^{1/2},
\end{equation}%
where $\xi $ is a parameter with value
\begin{equation}
\xi =\left( \frac{(D-1)^{2}\eta ^{2}}{16L^{2}}+\gamma m^{2}\right) ^{-1/2}.
\end{equation}%
We will show that $\xi $ is the correlation length of the boundary system
and that the correlation is short-ranged, in the later sections.

Around the boundary $z=0$, Eq. (\ref{U_solution_02}) could be expanded as%
\begin{equation}
\phi (z,k)=\phi _{0}(k)\left( z^{\frac{D}{2}-\nu }\left( 1+O(z)\right) +%
\frac{G(k)}{2\nu }z^{\frac{D}{2}+\nu }\left( 1+O(z)\right) \right) ,
\label{modes_03}
\end{equation}%
with $\phi _{0}(k)$ a $z$-independent prefactor and
\begin{equation}
G(k)=2\nu \frac{4^{-\nu }\Gamma (\alpha +\nu )\Gamma (-2\nu )}{\Gamma (2\nu
)\Gamma (\alpha -\nu )}\left( \frac{\beta }{L}\right) ^{2\nu }.
\end{equation}

In the regime $|\eta |\ll 1$, we have that $\frac{1}{2}-\alpha =O(|\eta |%
\frac{\xi }{8L})\ll 1$, i.e. $\alpha \approx \frac{1}{2}$.$~$Using the
Legendre duplication formula ${\Gamma (2z)=}$ $\frac{{2^{2z-1}{}}}{\sqrt{{%
\pi }}}{\Gamma (z)\Gamma (z+{\frac{1}{2}}),}$ we have the identity of Gamma
functions%
\begin{equation}
\frac{\Gamma \left( \nu +\frac{1}{2}\right) \Gamma (-2\nu )}{\Gamma (2\nu
)\Gamma \left( \frac{1}{2}-\nu \right) }=\frac{2^{-4\nu }\Gamma (-\nu )}{%
\Gamma (\nu )}.
\end{equation}%
Hence the response function $G(k)$ reads
\begin{equation}
G(k)=-2\nu \frac{2^{-2\nu }\Gamma (1-\nu )\left( k^{2}+\xi ^{-2}\right)
^{\nu }}{\Gamma (1+\nu )}.
\end{equation}

In the limit $\eta $ $\rightarrow 0,\gamma \rightarrow 0$ and hence $\frac{%
\beta z}{2L}\rightarrow 2kz$, by the Kummer's second transformation \cite%
{Abramowitz Stegun,Gradshteyn Ryzhik}, $U(a,2a,x)=\frac{e^{x/2}}{\sqrt{\pi }}%
x^{\frac{1}{2}-a}K_{a-\frac{1}{2}}\left( \frac{x}{2}\right) $. In this
limit, the above response function $G(k)$ reduces to $-2\nu \frac{2^{-2\nu
}\Gamma (1-\nu )}{\Gamma (1+\nu )}k^{2\nu }$, which is the result in the
maximally symmetric AdS geometry, see e.g. \cite%
{DHoker:2002nbb,Freedman:1998tz}.

\section{Bulk-to-bulk propagator and boundary-to-bulk propagator}
\renewcommand{\theequation}{4.\arabic{equation}} \setcounter{equation}{0}

Now we compute the bulk-to-bulk propagator and the boundary-to-bulk
propagator in the background (\ref{metric_01}) and (\ref{metric_02}). The
bulk-to-bulk propagator $G(x,z;x^{\prime },z^{\prime })$ in position space
is defined by
\begin{equation}
(\nabla ^{2}-m^{2})G(x,z;x^{\prime },z^{\prime })=-\frac{1}{\sqrt{|g|}}%
\delta ^{D}(x-x^{\prime })\delta (z-z^{\prime }),
\end{equation}%
where $x$ and $x^{\prime }$ denote $D$-dimensional position vectors, and $%
\nabla ^{2}={\frac{1}{\sqrt{|g|}}}\partial _{M}(\sqrt{|g|}g^{MN}\partial
_{N})$ is the Laplacian on the curved background. The bulk-to-bulk
propagator $\tilde{G}(k,z;z^{\prime })$ in momentum space satisfies%
\begin{equation}
\left( \partial _{z}^{2}+(D-1)\frac{a^{\prime }(z)}{a(z)}\partial
_{z}-k^{2}-a^{2}(z)m^{2}\right) \tilde{G}(k,z;z^{\prime })=-\frac{1}{%
a^{D-1}(z)}\delta (z-z^{\prime }).  \label{tildeG_eq}
\end{equation}%
We find a special solution of $\tilde{G}(k,z;z^{\prime })$ which could be
checked by substituting into it,
\begin{equation}
\tilde{G}(k,z;z^{\prime })=\frac{\theta (z-z^{\prime })\phi _{1}(z)\phi
_{2}(z^{\prime })+\theta (z^{\prime }-z)\phi _{1}(z^{\prime })\phi _{2}(z)}{%
a^{D-1}(z^{\prime })\left( \phi _{1}(z^{\prime })\phi _{2}^{\prime
}(z^{\prime })-\phi _{1}^{\prime }(z^{\prime })\phi _{2}(z^{\prime })\right)
},  \label{tildeG}
\end{equation}%
where $\phi _{1}(z)$ and $\phi _{2}(z)$ are two linearly independent
solutions of the Helmholtz equation \eqref{new_Helmholtz_02}. The
boundary-to-bulk propagator $\tilde{K}(z,k)$ in momentum space could be
derived by taking limit $z^{\prime }\rightarrow 0$ together with a
normalization factor, $\tilde{K}(z,k)=\lim_{z^{\prime }\rightarrow 0}2\nu
(z^{\prime })^{\frac{D}{2}-1-\nu }a^{D-1}(z^{\prime })\tilde{G}%
(k,z;z^{\prime })$. Thus the boundary-to-bulk propagator $K(z,x;x^{\prime })$
in position space could be gotten by Fourier transformation,
\begin{eqnarray}
&&%
\begin{split}
& K(z,x;x^{\prime })=\int \frac{\mathrm{d}^{D}\vec{k}}{(2\pi )^{D}}\tilde{K}%
(z,k)e^{i\vec{k}(\vec{x}-\vec{x}^{\prime })} \\
=& \frac{2^{-\nu +1}}{(2\pi )^{D/2}\Gamma (\nu )}\left( \frac{z}{\xi \sqrt{%
z^{2}+|x-x^{\prime }|^{2}}}\right) ^{\nu +\frac{D}{2}}e^{-\frac{(D-1)\eta z}{%
4L}}K_{\nu +\frac{D}{2}}\left( \xi ^{-1}\sqrt{z^{2}+|x-x^{\prime }|^{2}}%
\right) ,
\end{split}
\notag \\
&&  \label{boundary_to_bulk_}
\end{eqnarray}%
where $K_{\nu +\frac{D}{2}}\left( \cdot \right) $ is the modified Bessel
function of the second kind. The detailed derivations are in Appendix D.

In the $\xi ^{-1}\rightarrow 0$ limit, the above boundary-to-bulk propagator
(\ref{boundary_to_bulk_}) reduces to that of AdS case. For small $\xi ^{-1}$%
,
\begin{equation}
K_{\nu +\frac{D}{2}}\left( \xi ^{-1}\sqrt{z^{2}+|x-x^{\prime }|^{2}}\right) =%
\frac{{\Gamma ({\nu +\frac{D}{2}})}}{2}\left( \frac{\xi ^{-1}\sqrt{%
z^{2}+|x-x^{\prime }|^{2}}}{2}\right) ^{-\nu -\frac{D}{2}}(1+O(\xi ^{-1})),
\end{equation}%
consequently,
\begin{equation}
\lim_{\xi \rightarrow +\infty }K(z,x;x^{\prime })={C_{\Delta }}\left( \frac{z%
}{z^{2}+|x-x^{\prime }|^{2}}\right) ^{\Delta },
\end{equation}%
with $C_{\Delta }=\frac{\Gamma (\Delta )}{\pi ^{D/2}\Gamma (\nu )}$ and $%
\Delta =\frac{D}{2}+\nu $.

The boundary-to-boundary propagator $\beta (x,x^{\prime })$ could be
obtained by taking the limit $z\rightarrow 0$ together with a normalization
factor $z^{-\Delta }$, namely,
\begin{equation}
\beta (x,x^{\prime })=\lim_{z\rightarrow 0}z^{-\Delta }K(z,x;x^{\prime })=%
\frac{2^{-\nu +1}}{(2\pi )^{D/2}\Gamma (\nu )}\frac{K_{\nu +\frac{D}{2}%
}\left( \xi ^{-1}|x-x^{\prime }|\right) }{\left( \xi |x-x^{\prime }|\right)
^{\nu +\frac{D}{2}}}.  \label{boundary_01}
\end{equation}%
For large $|x-x^{\prime }|\gg \xi ,$
\begin{equation}
\beta (x,x^{\prime })\approx \frac{2^{-\nu +\frac{1}{2}}\sqrt{\pi }}{(2\pi
)^{D/2}\Gamma (\nu )}\frac{1}{\xi ^{2(\nu +\frac{D}{2})}}\exp \left[
-|x-x^{\prime }|/\xi +O(\ln (|x-x^{\prime }|/\xi ))\right] .
\label{boundary_large_separation_01}
\end{equation}%
For small $|x-x^{\prime }|\ll \xi ,$
\begin{equation}
\beta (x,x^{\prime })\approx \frac{\Gamma (\nu +\frac{D}{2})}{\pi
^{D/2}\Gamma (\nu )|x-x^{\prime }|^{2(\nu +\frac{D}{2})}}.
\end{equation}%
The Fourier transform of (\ref{boundary_01}) is%
\begin{eqnarray}
\tilde{\beta}(k) &=&\int \mathrm{d}^{D}x\beta (x,0)e^{-i\vec{k}\cdot \vec{x}}
\notag \\
&=&(2\pi )^{\frac{D}{2}}k^{1-\frac{D}{2}}\int_{0}^{\infty }\mathrm{d}xx^{%
\frac{D}{2}}J_{\frac{D}{2}-1}(kx)\beta (x,0)  \notag \\
&=&\frac{2^{-\nu +1}k^{1-\frac{D}{2}}}{\Gamma (\nu )\xi ^{\Delta }}%
\int_{0}^{\infty }\mathrm{d}x\,x^{\frac{D}{2}-\Delta }J_{\frac{D}{2}%
-1}(kx)K_{\Delta }(x/\xi ).
\end{eqnarray}%
The integral, however, could be worked out explicitly for $\Delta -\frac{D}{2%
}<0$,
\begin{equation}
\int_{0}^{\infty }\mathrm{d}x\,x^{\frac{D}{2}-\Delta }J_{\frac{D}{2}%
-1}(kx)K_{\Delta }(x/\xi )=2^{-\nu -1}k^{\frac{D}{2}-1}\xi ^{\Delta }{\Gamma
(-\nu )}{(k^{2}+\xi ^{-2})^{\Delta -\frac{D}{2}}}
\end{equation}%
which eventually gives
\begin{equation}
\tilde{\beta}(k)=\frac{2^{-2\nu }{\Gamma (-\nu )}}{\Gamma (\nu )}{(k^{2}+\xi
^{-2})^{\Delta -\frac{D}{2}}.}  \label{boundary_02}
\end{equation}%
For the case $\Delta -\frac{D}{2}\geq 0,$ we make a regularization with a
short-distance regulator, subtracting a regulator dependent piece, and then
the resulting integral is precisely (\ref{boundary_02}).

In the limit $\xi ^{-1}\rightarrow 0,$ the results here reduce to the
results on the maximally symmetric AdS geometry, see for example \cite%
{DHoker:2002nbb,Freedman:1998tz}.

\section{Derivation of short range correlation from gravity side}
\renewcommand{\theequation}{5.\arabic{equation}} \setcounter{equation}{0}
Here we derive the short-range correlation function of the boundary system
from the gravity computation. An integration by parts of the scalar field
action \eqref{scalar_action_AdS_01} gives
\begin{equation}
S_{\mathrm{reg}}\,=-\,{\frac{1}{2}}\int_{z\geq \epsilon }\mathrm{d}^{D}xdz%
\sqrt{|g|}[\phi (-\nabla ^{2}\,+\,m^{2})\,\phi \,]\,+{\frac{1}{2}}\int
\mathrm{d}^{D}x[\sqrt{|g|}g^{zz}\phi \partial _{z}\,\phi \,]\,|_{z=\epsilon
}\ .  \label{S_reg_01}
\end{equation}%
The $\epsilon $ is a UV regulator of the boundary system. The last term can
also be rewritten as ${\frac{1}{2}}\int \mathrm{d}^{D}x[\sqrt{|\gamma |}\phi
n^{\mu }\partial _{\mu }\,\phi \,]\,|_{z=\epsilon }$, with $\gamma _{\mu \nu
}=g_{MN}\frac{\partial y^{M}}{\partial x^{\mu }}\frac{\partial y^{N}}{%
\partial x^{\nu }}=a^{2}(z)\eta _{\mu \nu }$ the induced metric on the
boundary, and $n^{\mu }\partial _{\mu }=\frac{1}{a(z)}\partial _{z}$ where $%
n^{\mu }$ is the unit vector normal to the boundary. Near the boundary $%
\gamma _{\mu \nu }|_{z=\epsilon }=\frac{L^{2}}{\epsilon ^{2}}\eta _{\mu \nu
} $. \vspace{0.25cm}

The bulk field $\phi $ is the convolution of the boundary field $\phi _{0}$
and the boundary-to-bulk propagator,%
\begin{equation}
\phi (z,x)=\int \mathrm{d}^{D}x^{\prime }K(z,x;x^{\prime })\phi
_{0}(x^{\prime }),
\end{equation}%
and $\phi (z,x)|_{z=\epsilon }=z^{\frac{D}{2}-\nu }\phi _{0}(x)|_{z=\epsilon
}$.

The first term in the regularized action (\ref{S_reg_01}) is vanishing for
on-shell configurations. Hence the regularized on-shell action is
\begin{eqnarray}
S_{\mathrm{reg}}\, &=&\,\,{\frac{1}{2}}\,\,\int \mathrm{d}%
^{D}x[a^{D-1}(z)\phi \partial _{z}\,\phi \,]\,|_{z=\epsilon }  \notag \\
&=&\,{\frac{L^{D-1}}{2}}\int \mathrm{d}^{D}x_{1}\mathrm{d}^{D}x_{2}[\phi
_{0}(x_{1})\mathcal{A}(x_{1},x_{2})\phi _{0}(x_{2})\,].
\end{eqnarray}%
Here%
\begin{eqnarray}
\mathcal{A}(x_{1},x_{2}) &=&z^{-D+1}\int \mathrm{d}^{D}xK(z,x;x_{1})\partial
_{z}K(z,x;x_{2})|_{z=\epsilon }  \notag \\
&=&(D-\Delta )\epsilon ^{-\nu }\delta ^{D}(x_{1}-x_{2})+D\frac{2^{-\nu +1}}{%
(2\pi )^{D/2}\Gamma (\nu )}\frac{K_{\nu +\frac{D}{2}}\left( \xi
^{-1}|x_{1}-x_{2}|\right) }{\left( \xi |x_{1}-x_{2}|\right) ^{\nu +\frac{D}{2%
}}},  \notag \\
&&
\end{eqnarray}%
where we used the property (\ref{boundary_01}) of the boundary-to-bulk
propagator. In order to precisely cancel the first divergent term when
taking the limit $\epsilon \rightarrow 0$, it is inevitable to add the
counter-term%
\begin{eqnarray}
S_{\mathrm{ct}} &=&-\,\,{\frac{L^{D-1}}{2}}\,(D-\Delta )\,\int \mathrm{d}%
^{D}x[\sqrt{|\gamma |}\phi ^{2}\,]\,|_{z=\epsilon }  \notag \\
\, &=&{\frac{L^{D-1}}{2}}\int \mathrm{d}^{D}x_{1}\mathrm{d}^{D}x_{2}[\phi
_{0}(x_{1})\mathcal{A}_{\mathrm{ct}}(x_{1},x_{2})\phi _{0}(x_{2})\,],
\end{eqnarray}%
with
\begin{eqnarray}
\mathcal{A}_{\mathrm{ct}}(x_{1},x_{2}) &=&-(D-\Delta )\epsilon ^{-\nu
}\delta ^{D}(x_{1}-x_{2})-2(D-\Delta )\frac{2^{-\nu +1}}{(2\pi )^{D/2}\Gamma
(\nu )}\frac{K_{\nu +\frac{D}{2}}\left( \xi ^{-1}|x_{1}-x_{2}|\right) }{%
\left( \xi |x_{1}-x_{2}|\right) ^{\nu +\frac{D}{2}}}.  \notag \\
&&
\end{eqnarray}

The renormalized action $S_{\mathrm{ren}}\,=\,S_{\mathrm{reg}}+S_{\mathrm{ct}%
}\,$reads
\begin{equation}
S_{\mathrm{ren}}\,=\,{\frac{L^{D-1}}{2}}(2\Delta -D)\int \mathrm{d}^{D}x_{1}%
\mathrm{d}^{D}x_{2}[\phi _{0}(x_{1})\beta (x_{1},x_{2})\phi _{0}(x_{2})\,]\,,
\end{equation}%
where%
\begin{equation}
\beta (x_{1},x_{2})=\frac{2^{-\nu +1}}{(2\pi )^{D/2}\Gamma (\nu )}\frac{%
K_{\nu +\frac{D}{2}}\left( \xi ^{-1}|x_{1}-x_{2}|\right) }{\left( \xi
|x_{1}-x_{2}|\right) ^{\nu +\frac{D}{2}}}.  \label{boundary_03}
\end{equation}%
In the following calculation, we set the unit of energy such that $L=1$. The
dual operator $\mathcal{O}(x)$ in the boundary system is sourced by $\phi
_{0}(x)$, and the generating functional for the correlation function of $%
\mathcal{O}(x)~$is%
\begin{equation}
W[\phi _{0}]=\langle \exp \int \mathrm{d}^{D}x[\phi _{0}(x)\mathcal{O}%
(x)]\rangle =e^{S_{\mathrm{ren}}[\phi _{0}]}.
\end{equation}%
The vacua expectation value is then
\begin{eqnarray}
\langle \mathcal{O}(x)\rangle &=&{\frac{\delta }{\delta \phi _{0}(x)}}\,S_{%
\mathrm{ren}}[\phi _{0}]  \notag \\
&=&(2\Delta -D)\int \mathrm{d}^{D}x^{\prime }[\phi _{0}(x^{\prime })\beta
(x,x^{\prime })]\,=\,(2\Delta -D)\phi _{1}(x).  \label{relation_}
\end{eqnarray}%
The two-point correlation function is then
\begin{eqnarray}
\langle \mathcal{O}(x)\mathcal{O}(x^{\prime })\rangle &=&{\frac{\delta }{%
\delta \phi _{0}(x)}\frac{\delta }{\delta \phi _{0}(x^{\prime })}}\,S_{%
\mathrm{ren}}[\phi _{0}]  \notag \\
&=&\,(2\Delta -D)\beta (x,x^{\prime }).  \label{two_point_}
\end{eqnarray}%
Here $\beta (x,x^{\prime })$ is (\ref{boundary_03}) which is short-range
correlated, and in the large separation $|x-x^{\prime }|$ it takes the form
of (\ref{boundary_large_separation_01}).

Via the boundary-to-bulk propagator,
\begin{equation}
\phi (z,x)=\int \mathrm{d}^{D}x^{\prime }K(z,x;x^{\prime })\phi
_{0}(x^{\prime }).
\end{equation}%
Performing Fourier transform,%
\begin{equation}
\phi (z,k)=\tilde{K}(z,k)\phi _{0}(k).
\end{equation}%
Near the boundary, $\phi (z,x)|_{z=\epsilon }=z^{\frac{D}{2}-\nu }\phi
_{0}(x)|_{z=\epsilon }~$, and hence their Fourier transforms satisfy $\phi
(z,k)|_{z=\epsilon }=z^{\frac{D}{2}-\nu }\phi _{0}(k)|_{z=\epsilon }~$. The
field near the boundary is $\phi (z,k)|_{z=\epsilon }=[\phi
_{0}(k)u_{0}(z)+\phi _{1}(k)u_{1}(z)]|_{z=\epsilon }$ where the two terms
are source-mode and vev-mode respectively. Around $z=0$, performing the
expansion of the propagator,
\begin{eqnarray}
\tilde{K}(z,k)|_{z=\epsilon } &=&\left. [z^{\frac{D}{2}-\nu }-\frac{2^{-2\nu
}\Gamma (1-\nu )}{\Gamma (1+\nu )}(k^{2}+\xi ^{-2})^{\Delta -\frac{D}{2}}z^{%
\frac{D}{2}+\nu }]\right\vert _{z=\epsilon }  \notag \\
&=&[z^{\frac{D}{2}-\nu }+\frac{G(k)}{2\nu }z^{\frac{D}{2}+\nu
}]|_{z=\epsilon }.
\end{eqnarray}%
Hence we see that $\phi _{1}(k)=\frac{G(k)}{2\nu }\phi _{0}(k)$. We are
still in an asymptotically AdS space, although we are not in a maximally
symmetric AdS geometry. Hence $G(k)$ is the response of the vev$~\phi _{1}(k)
$ to the source $\phi _{0}(k)$. Performing the Fourier transform of (\ref%
{relation_}),%
\begin{equation}
\phi _{1}(k)=\tilde{\beta}(k)\phi _{0}(k),
\end{equation}%
so we see that $G(k)=2\nu \tilde{\beta}(k)$.

Performing Fourier transform of the fields, we hence have in the momentum
space,
\begin{equation}
\langle \mathcal{O}(k)\mathcal{O}(-k)\rangle =(2\Delta -D)\tilde{\beta}%
(k)=-2\nu \frac{2^{-2\nu }\Gamma (1-\nu )}{\Gamma (1+\nu )}(k^{2}+\xi
^{-2})^{\Delta -\frac{D}{2}}.
\end{equation}%
This is the Fourier transform of (\ref{two_point_}).

In the limit $\xi ^{-1}\rightarrow 0$, the formula between the vev and
source reduces to the canonical result in the maximally symmetric AdS
geometry, e.g. \cite%
{DHoker:2002nbb,Freedman:1998tz,Banks:1998dd,Balasubramanian:1998de}.

\section{Relation to correlation length}
\renewcommand{\theequation}{6.\arabic{equation}} \setcounter{equation}{0}

Here we analyze the implication of our gravity computation to the
short-range correlation and the correlation length $\xi $. We look at
features of the two-point function in position space. For small $%
|x-x^{\prime }|\ll \xi $,
\begin{equation}
\langle \mathcal{O}(x)\mathcal{O}(x^{\prime })\rangle \approx 2\nu C_{\Delta
}\frac{1}{|x-x^{\prime }|^{2\Delta }}.  \label{small_separation_}
\end{equation}%
The operator $\mathcal{O}(x)$ has a UV scaling dimension of $\Delta .$

For large $|x-x^{\prime }|\gg \xi ,$
\begin{equation}
\langle \mathcal{O}(x)\mathcal{O}(x^{\prime })\rangle \approx C\frac{1}{\xi
^{2\Delta }}\exp \left[ -|x-x^{\prime }|/\xi +O(\ln (|x-x^{\prime }|/\xi ))%
\right] ,
\end{equation}%
where $C=2\nu \frac{2^{-\nu +\frac{1}{2}}\sqrt{\pi }}{(2\pi )^{D/2}\Gamma
(\nu )}$. We may rescale the fields such that $\tilde{\mathcal{O}}=\xi
^{\Delta }\mathcal{O}$, hence in large separation,%
\begin{equation}
\langle \tilde{\mathcal{O}}(x)\tilde{\mathcal{O}}(x^{\prime })\rangle
\approx C\exp \left[ -|x-x^{\prime }|/\xi +O(\ln (|x-x^{\prime }|/\xi ))%
\right] .  \label{large_separation_}
\end{equation}%
Eq. \eqref{large_separation_} shows that the system has a short-range
correlation function around the ground state. The leading behavior of this
short-range correlation function is an exponential decay $e^{-|x-x^{\prime
}|/\xi }$ with respect to the separation $|x-x^{\prime }|$ with a
correlation length $\xi $.

The inverse correlation length $\xi ^{-1}$ is related to the inverse radius
size $z_{0}^{-1}$ in the holographic dimension. We have that $\beta
=4L(k^{2}+\xi ^{-2})^{1/2}$. The\ $\xi ^{-1}$ is the inverse correlation
length, which is an energy scale in the infrared regime of the boundary
system. In the dual geometric side, $z_{0}^{-1}$ is an infrared energy scale
for the boundary system. In the model here, as derived from the response
function, we have
\begin{equation}
\xi ^{-1}=\sqrt{\frac{(D-1)^{2}}{16L^{2}}\eta (z_{0})^{2}+\gamma (z_{0})m^{2}%
}.
\end{equation}%
If$~\eta (z_{0})=\bar{\eta}(\frac{L}{z_{0}})^{a}$,$~\gamma (z_{0})=\bar{%
\gamma}(\frac{L^{2}}{z_{0}^{2}})^{a}$ where $\bar{\eta},\bar{\gamma}$ are
dimensionless numbers and $a$ is any non-negative number, then
\begin{equation}
\xi ^{-1}=c(\nu )L^{a-1}z_{0}^{-a},  \label{correlation_length_04}
\end{equation}%
where $c(\nu )=\sqrt{\frac{(D-1)^{2}}{16}\bar{\eta}^{2}+\bar{\gamma}%
m^{2}L^{2}}~$is a model-dependent factor.

Hence the correlation length $\xi ~$of the boundary system is mapped to a
geometric radius size $z_{0}$ in the dual gravity description. For example,
we expect $\xi ^{-1}\propto z_{0}^{-a}$, such as in Eq. (\ref%
{correlation_length_04}). The dual boundary system has a short-range
correlation function which is an exponential decay in large separation,
evaluated near its ground state. This is a nice feature from our gravity
description. On the other hand, short-range correlation functions are very
interesting and can be exhibited in many systems, such as in quantum
magnetic systems, in quantum vortex models, and in confining gauge theories.

\section{Discussion}
\label{sec_discussion}

We constructed a class of backgrounds with a warp factor $a^{2}(z)$ and
anti-de Sitter asymptotics, which are dual to boundary systems that have a
ground state with a short-range two-point correlation function. We produced
these metrics in matter coupled gravity using scalar coupled to gravity as
an example, and our scalar profile deforms the background. This is a
two-parameter class of geometries for short-range correlations on the
boundary. Our solutions hold for a general $D$.

We computed the bulk-to-boundary Green's function for probe scalar fields in
these backgrounds, obtained from our solutions of the fields on these
backgrounds by means of confluent hypergeometric functions. Using them, the
two-point correlation function for operators in the boundary theory which
couple to the boundary values of the probe scalar field, were calculated.
This leads to a short-range two-point function with a correlation length $%
\xi $. Explicit analytical expressions of the correlation functions with the
short-range correlation and a correlation length were obtained. Our
analytical expressions may be useful for further aspects of the
correspondence in this set-up. We also obtained the relation between
source-mode and vev-mode in these backgrounds, which, in the limit when $\xi
^{-1}$ goes to zero, reduce to the canonical results in the maximally
symmetric AdS geometry.

The operator aforementioned has a naive scaling dimension in the UV, since
our set-up is still in asymptotically AdS geometry. In the IR, its two-point
correlation function will decay exponentially with the separation, with a
characteristic length scale the correlation length $\xi $. We computed this
correlation length from the gravity side, using gauge/gravity
correspondence. Hence our gravity results predict that the boundary system
has a short-range correlation with a correlation length derived from gravity.

The inverse correlation length in this paper is not due to a thermal nature.
The topology of the metrics does not have a temporal circle whose perimeter
would represent the inverse finite temperature. The correlation function
here is evaluated near the ground state. Our correlation functions are
similar to the short-range two-point correlation functions in a massive
quantum field theory at zero temperature or a quantum many-body system near
a ground state with a spectral gap. However, finite temperature deformation
of our backgrounds is possible, which might lead to a finite temperature
deformation of our correlation length.

We analyzed the implication of our gravity computation of the short-range
correlation and the correlation length $\xi $. The boundary systems thus
have a short-range correlation near the ground state, and this feature is
similar to that of certain strongly correlated systems, such as those in
symmetry breaking phases. On the other hand, there are many condensed matter
systems which have short-range correlation functions evaluated near the
ground states, for example, the non-linear sigma model \cite{NLSM}, the
nearest-neighbor resonating-valence-bond state \cite{NNRVB}, quantum vortex
models \cite{votex}, and many other strongly correlated models such as
quantum XY model in the ordered phase \cite{XY}.

The feature that the boundary system has a short-range correlation near the
ground state, is also similar to that of confining gauge theories. Our
set-up may also be related to Improved Holographic QCD models \cite%
{Gursoy:2010fj}, where features of confinement and discrete spectrum, among
other things, are produced in the gravity dual. In that context, they have
used Einstein gravity coupled with dilaton. These above aspects are all very
interesting for further investigations.

\section*{Acknowledgments}

We would like to thank J. P. Shock, N. Su, Y.-W. Sun, and P. Zhao for
communications or discussions. The work was supported in part by Yau
Mathematical Sciences Center and by Tsinghua University.

\appendix

\section{Matter coupled gravity}
\label{appendix_special_case} \renewcommand{\theequation}{A.%
\arabic{equation}} \setcounter{equation}{0} \renewcommand{\thethm}{A.%
\arabic{thm}} \setcounter{thm}{0} \renewcommand{\theprop}{A.\arabic{prop}} %
\setcounter{prop}{0}

The gravity dual is on the manifold $M^{D+1}$, with the metric of the form
\begin{equation}
ds^{2}=a^{2}(z)(\mathrm{d}z^{2}+\eta _{\mu \nu }\mathrm{d}x^{\mu }\mathrm{d}%
x^{\nu }),
\end{equation}%
where $\mu =0,\cdots ,D-1$. We consider the background with the warp factor $%
a^{2}(z)$ in our case, to be dual to a boundary system that has a ground
state with a short-range two-point correlation function.

By choosing appropriate matter and its potential, we can obtain the geometry
with the warp factor $a^{2}(z)$. These warp factors can be obtained in
matter coupled gravity, with appropriate matter sources. The matter sector
does not have to be unique. One can choose matter sectors to be scalar
field, axion-scalar field, vector field, or fermionic field. These matter
fields back-react to the gravity sector, and thus deform the background
geometry.

Consider that the back-reacting matter is modeled by a scalar field $\varphi
~$with a potential $V(\varphi )$. The action of the matter with gravity is%
\begin{equation}
S=\frac{1}{2\kappa ^{2}}\int \mathrm{d}^{D+1}y\sqrt{-g}\left[ R-{\frac{1}{2}}%
\partial ^{M}\varphi \partial _{M}\varphi -V(\varphi )\right] ,
\label{matter_gravity_01}
\end{equation}%
where $\kappa $ is the gravitational coupling constant, and $M=0,\cdots ,D.$

The geometry is a small deformation of AdS$_{D+1}$ spacetime. We consider
the warp factor to be
\begin{equation}
a^{2}(z)=\frac{L^{2}}{z^{2}}+\frac{\eta L}{z}+\gamma +O(\frac{z}{L}),
\end{equation}%
where $\eta =\eta (z_{0}),\gamma =\gamma (z_{0})$ are functions of $z_{0}$.

The equations of motion obtained from (\ref{matter_gravity_01}) are%
\begin{equation}
R_{MN}-{\frac{1}{2}}g_{MN}\left( R-{\frac{1}{2}}\partial ^{I}\varphi
\partial _{I}\varphi -V(\varphi )\right) -{\frac{1}{2}}\partial _{M}\varphi
\partial _{N}\varphi =0.  \label{eom_01_1}
\end{equation}%
\begin{equation}
\nabla ^{M}\nabla _{M}\varphi -\frac{\mathrm{d}V(\varphi )}{\mathrm{d}%
\varphi }=0.  \label{eom_01_2}
\end{equation}

The equations of the motion simplified from the above (\ref{eom_01_1}), when
we consider the solutions invariant under the isometry of the boundary are
\begin{equation}
(D-1)\left( \frac{a^{\prime \prime }(z)}{a(z)}-2\left( \frac{a^{\prime }(z)}{%
a(z)}\right) ^{2}\right) +\frac{1}{2}\left( \varphi ^{\prime }(z)\right)
^{2}=0.  \label{eom_02_1}
\end{equation}%
\begin{equation}
(D-1)\left( \frac{a^{\prime \prime }(z)}{a(z)}\right) +(D-1)(D-2)\left(
\frac{a^{\prime }(z)}{a(z)}\right) ^{2}+a^{2}(z)V(\varphi )=0.
\label{eom_02_2}
\end{equation}%
There is a third equation simplified from (\ref{eom_01_2}),
\begin{equation}
a^{-2}(z)\left( \varphi ^{\prime \prime }(z)+(D-1)\frac{a^{\prime }(z)}{a(z)}%
\varphi ^{\prime }(z)\right) -\frac{\mathrm{d}V(\varphi )}{\mathrm{d}\varphi
}=0.  \label{eom_02_3}
\end{equation}

With appropriate potential, the metric can be generated. The solutions
depend on not only the potential $V(\varphi )$ but also the profile $\varphi
(z)$. From (\ref{eom_02_1}), the derivative of the $\varphi (z)$ is
\begin{equation}
\varphi ^{\prime }(z)=\pm \sqrt{-\frac{(D-1)L\left( 4\eta L^{2}+Lz\left(
12\gamma +\eta ^{2}\right) +4\gamma \eta z^{2}\right) }{2z\left( L^{2}+\eta
Lz+\gamma z^{2}\right) ^{2}}}.
\end{equation}%
To make sure $\varphi ^{\prime }(z)$ is real,
\begin{equation}
4\eta L^{2}+Lz\left( 12\gamma +\eta ^{2}\right) +4\gamma \eta z^{2}<0,\qquad
\forall z\geq 0.
\end{equation}%
Take limit $z\rightarrow 0^{+}$, we could get $\eta \leq 0$. So $\eta $ must
be non-positive and $\gamma $ is non-negative.

We make a Taylor expansion with respect to $z$ up to the first three orders.
Hence, in a parametric representation, for%
\begin{equation}
a^{2}(z)=\frac{L^{2}}{z^{2}}+\frac{\eta L}{z}+\gamma +O(\frac{z}{L}),
\label{metric_01_}
\end{equation}%
to the first three orders in $z$, the scalar and its potential are
\begin{equation}
V=-\frac{D(D-1)}{L^{2}}+(D-1)(2D-1)\frac{\eta z}{L^{3}}+(D-1)(12D\gamma
-12\gamma -13D\eta ^{2}+11\eta ^{2})\frac{z^{2}}{4L^{2}},  \label{v_01}
\end{equation}%
\begin{equation}
\varphi =\varphi _{0}\mp \frac{1}{6}\sqrt{\frac{(D-1)z}{-2\eta L}}\left[
24\eta +(12\gamma -7\eta ^{2})\frac{z}{L}\right] .  \label{profile_01}
\end{equation}%
Note that (\ref{v_01})--(\ref{profile_01}) is a parametric form of the
potential $V(\varphi )$, through the variable $z$. The ratio $\gamma /\eta
^{2}$ can take general values. The meaning of $\varphi _{0}$ is that it is
the value of $\varphi $ at $z=0$ with $V(\varphi _{0})=-\frac{D(D-1)}{L^{2}}$%
. The AdS case is\ $V=-\frac{D(D-1)}{L^{2}}\mathrm{\ }$and $\varphi =\varphi
_{0}$, with $\eta =0$ and $\gamma =0.$

In particular, we also have an interesting special solution%
\begin{equation}
a(z)=L(\frac{1}{z}-\frac{1}{z+2z_{0}}),  \label{metric_02_}
\end{equation}%
for $z\geq 0$ and $2z_{0}>0,$ and we have that
\begin{equation}
V(\varphi )=-\frac{(D-1)}{8L^{2}}\left( (2D-1)e^{\frac{\varphi -\varphi _{0}%
}{\sqrt{D-1}}}+(2D-1)e^{\frac{\varphi _{0}-\varphi }{\sqrt{D-1}}%
}+2(2D+1)\right) ,  \label{v_02}
\end{equation}%
\begin{equation}
\varphi =\varphi _{0}\pm 4\sqrt{D-1}\log \left( \sqrt{\frac{z}{2z_{0}}+1}+%
\sqrt{\frac{z}{2z_{0}}}\right) .  \label{profile_02}
\end{equation}%
And the above solution (\ref{metric_02_})--(\ref{profile_02}) is an exact
solution. The above solution can reduce to AdS case in the $%
z_{0}^{-1}\rightarrow 0$ limit.

The case $a(z)=L/z-L/(z+2z_{0})$ for $z\geq 0$ and $2z_{0}>0$ corresponds to
$\eta =-Lz_{0}^{-1},\gamma =\frac{3}{4}L^{2}z_{0}^{-2}$. The inverse
function of (\ref{profile_02}) is

\begin{equation}
z=\frac{z_{0}}{2}e^{-\frac{\varphi _{0}}{2\sqrt{D-1}}-\frac{\varphi }{2\sqrt{%
D-1}}}\left( e^{\frac{\varphi }{2\sqrt{D-1}}}-e^{\frac{\varphi _{0}}{2\sqrt{%
D-1}}}{}\right) ^{2}.
\end{equation}%
In this variable,

\begin{equation}
V=-\frac{(D-1)\left( (2D-1)z^{2}+2(2D-1)zz_{0}+2Dz_{0}^{2}\right) }{%
2L^{2}z_{0}^{2}}  \label{v_02_}
\end{equation}%
exactly. The profile (\ref{profile_02}) can be expanded as
\begin{equation}
\varphi (z)=\varphi _{0}\mp \frac{(z-12z_{0})\sqrt{(D-1)zz_{0}}}{3\sqrt{2}%
z_{0}^{2}}.
\end{equation}%
It is exactly Eq.(\ref{profile_01}) with $\eta =-L/z_{0}$ and $\gamma
=3L^{2}/(4z_{0}^{2})$. Inputting these values in Eq. (\ref{v_01}) gives (\ref%
{v_02_}) exactly. This is an example when $\gamma /\eta ^{2}=3/4$ which is
of order 1. On the other hand, we can also have other examples when$~\gamma
/\eta ^{2}$ take more general values including when $\gamma /\eta ^{2}$ is
large, from the more general solutions in (\ref{metric_01_})--(\ref%
{profile_01}). Our solutions hold for a general $D$.

\section{Relation to holographic renormalization}
\renewcommand{\theequation}{B.\arabic{equation}} \setcounter{equation}{0} %
\renewcommand{\thethm}{B.\arabic{thm}} \setcounter{thm}{0} %
\renewcommand{\theprop}{B.\arabic{prop}} \setcounter{prop}{0}

We are in matter coupled gravity. The $z$ is a holographic radial direction.
Our metrics are relevant for holographic normalization schemes, e.g. \cite%
{Skenderis:2002wp,Papadimitriou:2004ap}. Here we describe relations between
our ansatz and that used in the holographic renormalization.

The metric ansatz is
\begin{equation}
\mathrm{d}s^{2}=a^{2}(z)\left( \mathrm{d}z^{2}+(\eta _{\mu \nu }+h_{\mu \nu
})\mathrm{d}x^{\mu }\mathrm{d}x^{\nu })\right) ,  \label{metric-03}
\end{equation}%
where $\eta _{\mu \nu }$ is the background metric of the boundary and $%
h_{\mu \nu }$ is its fluctuation. We can write this metric in the form used
in holographic renormalization schemes \cite%
{Skenderis:2002wp,Papadimitriou:2004ap}
\begin{equation}
\mathrm{d}s^{2}=L^{2}\left( \frac{\mathrm{d}r^{2}}{r^{2}}+\frac{1}{r^{2}}%
g_{\mu \nu }\mathrm{d}x^{\mu }\mathrm{d}x^{\nu }\right) .  \label{metric-04}
\end{equation}%
In the above, $r$ is a holographic radial direction, which is related to $z$
by a change of coordinates. In particular, if the background is AdS, then $%
r=z$.

The change of coordinates between (\ref{metric-03}) and (\ref{metric-04}) is
\begin{equation}
\log r=C+\frac{1}{L}\int a(z)\mathrm{d}z,\qquad g_{\mu \nu }=r^{2}\frac{%
a^{2}(z)}{L^{2}}(\eta _{\mu \nu }+h_{\mu \nu }).
\end{equation}

For $a(z)=L/z-L/(z+2z_{0})$, the integral gives
\begin{equation}
r=\frac{2z_{0}z}{z+2z_{0}}={z}-\frac{z^{2}}{2z_{0}}+\frac{z^{3}}{4z_{0}^{2}}%
+O(z^{4}),
\end{equation}%
\begin{equation}
z=\frac{2rz_{0}}{2z_{0}-r}=r+\frac{r^{2}}{2z_{0}}+\frac{r^{3}}{4z_{0}^{2}}%
+O(r^{4}),
\end{equation}

\begin{equation}
g_{\mu \nu }=\frac{(r-2z_{0})^{4}}{16z_{0}^{4}}(\eta _{\mu \nu }+h_{\mu \nu
})=\left( 1-\frac{2r}{z_{0}}+\frac{3r^{2}}{2z_{0}^{2}}+O\left( \frac{r^{3}}{%
L^{3}}\right) \right) (\eta _{\mu \nu }+h_{\mu \nu }).  \label{exp_01}
\end{equation}%
We are in matter coupled gravity, hence we keep both even and odd powers of $%
r$, in the expansion of $g_{\mu \nu }$.

For $a^{2}(z)=L^{2}/z^{2}+\eta L/z+\gamma +O(z/L),$
\begin{equation}
a(z)=\frac{L}{z}+\frac{\eta }{2}-\frac{\eta ^{2}-4\gamma }{8L}z+O\left(
\frac{z^{2}}{L^{2}}\right) ,
\end{equation}%
\begin{equation}
r=z+\frac{\eta }{2L}z^{2}+\frac{\eta ^{2}+4\gamma }{16L^{2}}z^{3}+O(z^{4}),
\end{equation}%
\begin{equation}
z=r-\frac{\eta }{2L}r^{2}+\frac{7\eta ^{2}-4\gamma }{16L^{2}}r^{3}+O(r^{4}),
\end{equation}%
\begin{equation}
g_{\mu \nu }=\left( 1+\frac{2\eta }{L}r+\frac{3(\eta ^{2}+4\gamma )}{8L^{2}}%
r^{2}+O\left( \frac{r^{3}}{L^{3}}\right) \right) (\eta _{\mu \nu }+h_{\mu
\nu }).  \label{exp_02}
\end{equation}%
The expansion of the first case (\ref{exp_01}) is a special case of the
above expansion (\ref{exp_02}) by substituting $\eta =-L/z_{0}$ and $\gamma
=3L^{2}/(4z_{0}^{2})$.

\section{Detailed derivation of the basis solutions}
\renewcommand{\theequation}{C.\arabic{equation}} \setcounter{equation}{0} %
\renewcommand{\thethm}{C.\arabic{thm}} \setcounter{thm}{0} %
\renewcommand{\theprop}{C.\arabic{prop}} \setcounter{prop}{0}

The Helmholtz equation (\ref{Helmholtz}) gives the following form
\begin{eqnarray}
&&\left( \left( k^{2}+m^{2}\gamma (z_{0})\right) +\frac{m^{2}L^{2}}{z^{2}}%
+m^{2}\eta (z_{0})\frac{L}{z}\right.   \notag \\
&&\left. -(D-1)\left( -\frac{1}{z}+\frac{\eta (z_{0})}{2L}+\frac{z}{L^{2}}%
\left( \gamma (z_{0})-\frac{\eta (z_{0})^{2}}{2}\right) \right) \partial
_{z}-\partial _{z}^{2}\right) \phi (z,k)=0.  \label{new_Helmholtz_02_ap}
\end{eqnarray}%
This equation has exact solutions if the linear term in the first-order
derivative term is neglected. The two linearly independent basis solutions
are by means of the confluent hypergeometric functions of the second kind $%
U(a,b,x)$ and of the first kind $_{1}F_{1}(a,b,x)$, respectively, namely
\begin{eqnarray}
\phi (z,k) &=&z^{\frac{D}{2}+\nu }e^{-\frac{(\beta +(D-1)\eta )z}{4L}}\left[
c_{(1)}U\left( \alpha +\nu ,2\nu +1,\frac{\beta z}{2L}\right)
+c_{(2)}\,_{1}F_{1}\left( \alpha +\nu ,2\nu +1,\frac{\beta z}{2L}\right) %
\right]   \notag \\
&=&c_{(1)}h_{1}(z,k)+c_{(2)}h_{2}(z,k).  \label{solution basis}
\end{eqnarray}%
The second equal sign gives the definition of $h_{1}(z,k)$ and $h_{2}(z,k)$,
and $c_{(1)},c_{(2)}$ are two coefficients. Here $\nu =\sqrt{\frac{D^{2}}{4}%
+m^{2}L^{2}}$, and we have $\alpha =\frac{1}{2}-\frac{\eta }{2\beta }\left[
(D-1)^{2}-4m^{2}L^{2}\right] $, $\beta =4L(k^{2}+\xi ^{-2})^{1/2}$, where $%
\xi $ is a parameter,
\begin{equation}
\xi =\left( \frac{(D-1)^{2}\eta ^{2}}{16L^{2}}+\gamma m^{2}\right) ^{-1/2}.
\end{equation}

The definition of the confluent hypergeometric functions are \cite%
{Abramowitz Stegun,Gradshteyn Ryzhik}
\begin{equation}
_{1}F_{1}(a,b,x)=\sum_{n=0}^{\infty }\frac{\Gamma (a+n)/\Gamma (a)\,x^{n}}{%
\Gamma (b+n)/\Gamma (b)~n!},
\end{equation}%
\begin{equation}
U(a,b,x)=\frac{\Gamma (1-b)}{\Gamma (a+1-b)}\,_{1}F_{1}(a,b,x)+\frac{\Gamma
(b-1)}{\Gamma (a)}x^{1-b}\,_{1}F_{1}(a+1-b,2-b,x).
\end{equation}%
The definition in the second line only works for non-integer $b$, but it can
be extended to any integers $b$ by continuity. The Kummer's transformations
\begin{equation}
U(a,b,x)=x^{1-b}U(a-b+1,2-b,x)
\end{equation}%
can be used to verify the $\nu \leftrightarrow -\nu $ symmetry of the
Helmholtz equation.

Around $z=0$, $h_{2}(z,k)=z^{\frac{D}{2}+\nu }\left( 1+O(z)\right) $, which
does not contain a $z^{\frac{D}{2}-\nu }$ mode. Whereas, $h_{1}(z,k)$ is
vanishing when $z$ goes to infinity, and its expansion around $z=0$ contains
both the $z^{\frac{D}{2}+\nu }$ mode and the $z^{\frac{D}{2}-\nu }$ mode.

Consider this special solution,
\begin{equation}
\phi (z,k)=z^{\frac{D}{2}+\nu }e^{-\frac{(\beta +(D-1)\eta )z}{4L}}U\left(
\alpha +\nu ,2\nu +1,\frac{\beta z}{2L}\right) .  \label{U_solution_02_ap}
\end{equation}%
Around the boundary $z=0$, Eq. (\ref{U_solution_02_ap}) could be expanded as
\begin{equation}
\phi (z,k)=z^{\frac{D}{2}-\nu }\frac{2^{2\nu }\Gamma (2\nu )}{\Gamma (\alpha
+\nu )}\left( \frac{\beta }{L}\right) ^{-2\nu }\left( 1+O(z)\right) +z^{%
\frac{D}{2}+\nu }\frac{\Gamma (-2\nu )}{\Gamma (\alpha -\nu )}\left(
1+O(z)\right) .
\end{equation}%
Hence it can be rewritten in the form
\begin{equation}
\phi (z,k)=\phi _{0}(k)\left( z^{\frac{D}{2}-\nu }(1+O(z))+\frac{G(k)}{2\nu }%
z^{\frac{D}{2}+\nu }(1+O(z))\right) ,  \label{modes_03_ap}
\end{equation}%
with $\phi _{0}(k)$ a $z$-independent prefactor and
\begin{equation}
G(k)=2\nu \frac{4^{-\nu }\Gamma (\alpha +\nu )\Gamma (-2\nu )}{\Gamma (2\nu
)\Gamma (\alpha -\nu )}\left( \frac{\beta }{L}\right) ^{2\nu }.
\end{equation}%
The pieces $O(z)$ in the source mode and in the vev mode in (\ref%
{modes_03_ap}) have the following property near the boundary, $\underset{%
\epsilon \rightarrow 0}{\mathrm{lim}}$ $z^{\frac{D}{2}-\nu }\left(
1+O(z)\right) |_{z=\epsilon }=\epsilon ^{\frac{D}{2}-\nu }$.

In the regime $|\eta |\ll 1$, we have that $\frac{1}{2}-\alpha =O(|\eta |%
\frac{\xi }{8L})\ll 1$, i.e. $\alpha \approx \frac{1}{2}.~$Using the
Legendre duplication formula ${\Gamma (2z)=}$ $\frac{{2^{2z-1}{}}}{\sqrt{{%
\pi }}}{\Gamma (z)\Gamma (z+{\frac{1}{2}})}${, }the response function $G(k)$
reads
\begin{equation}
G(k)=-2\nu \frac{2^{-2\nu }\Gamma (1-\nu )\left( k^{2}+\xi ^{-2}\right)
^{\nu }}{\Gamma (1+\nu )}.
\end{equation}

Consider the limit $\eta $ $\rightarrow 0,\gamma \rightarrow 0$ and hence $%
\frac{\beta z}{2L}\rightarrow 2kz$. By the Kummer's second transformation
\cite{Abramowitz Stegun,Gradshteyn Ryzhik},
\begin{equation}
U(a,2a,x)=\frac{e^{x/2}}{\sqrt{\pi }}x^{\frac{1}{2}-a}K_{a-\frac{1}{2}%
}\left( \frac{x}{2}\right) ,
\end{equation}%
\begin{equation}
\,_{1}F_{1}(a,2a,x)=4^{a-\frac{1}{2}}e^{x/2}x^{\frac{1}{2}-a}\Gamma (a+\frac{%
1}{2})I_{a-\frac{1}{2}}\left( \frac{x}{2}\right) ,
\end{equation}%
the above solutions reduce to
\begin{equation}
h_{1}(z,k)=e^{-kz}z^{\frac{D}{2}+\nu }U\left( \frac{1}{2}+\nu ,2\nu
+1,2kz\right) =\frac{(2k)^{-\nu }}{\sqrt{\pi }}z^{D/2}K_{\nu }(kz),
\end{equation}%
\begin{equation}
h_{2}(z,k)=e^{-kz}z^{\frac{D}{2}+\nu }\,_{1}F_{1}\left( \frac{1}{2}+\nu
,2\nu +1,2kz\right) =\left( \frac{2}{k}\right) ^{\nu }\Gamma (\nu
+1)z^{D/2}I_{\nu }(kz),
\end{equation}%
and $G(k)$ reduces to $-2\nu \frac{2^{-2\nu }\Gamma (1-\nu )}{\Gamma (1+\nu )%
}k^{2\nu }$, which are the results in the maximally symmetric AdS geometry,
see e.g. \cite{DHoker:2002nbb,Freedman:1998tz}.

\section{Detailed derivation of the bulk-to-bulk and boundary-to-bulk
propagators}
\renewcommand{\theequation}{D.\arabic{equation}} \setcounter{equation}{0} %
\renewcommand{\thethm}{D.\arabic{thm}} \setcounter{thm}{0} %
\renewcommand{\theprop}{D.\arabic{prop}} \setcounter{prop}{0}

Here we present details of the derivation in Section 4. The bulk-to-bulk
propagator $\tilde{G}(k,z;z^{\prime })$ could be written as
\begin{equation}
\tilde{G}(k,z;z^{\prime })=\frac{\theta (z-z^{\prime })\phi _{1}(z)\phi
_{2}(z^{\prime })+\theta (z^{\prime }-z)\phi _{1}(z^{\prime })\phi _{2}(z)}{%
a^{D-1}(z^{\prime })\left( \phi _{1}(z^{\prime })\phi _{2}^{\prime
}(z^{\prime })-\phi _{1}^{\prime }(z^{\prime })\phi _{2}(z^{\prime })\right)
},  \label{tildeG_ap}
\end{equation}%
where $\phi _{1}(z)$ and $\phi _{2}(z)$ are two linearly independent
solutions of the Helmholtz equation \eqref{new_Helmholtz_02}.

As derived in Section 3, we let
\begin{equation}
\phi _{1}(z)=z^{\frac{D}{2}+\nu }e^{-\frac{z(\beta +(D-1)\eta )}{4L}}U\left(
\alpha +\nu ,2\nu +1,\frac{\beta z}{2L}\right) ,  \label{basis_01_1}
\end{equation}%
\begin{equation}
\phi _{2}(z)=z^{\frac{D}{2}+\nu }e^{-\frac{z(\beta +(D-1)\eta )}{4L}%
}{}_{1}F_{1}\left( \alpha +\nu ,2\nu +1,\frac{\beta z}{2L}\right) .
\label{basis_01_2}
\end{equation}%
Consequently the bulk-to-bulk propagator can be written as%
\begin{equation}
\begin{split}
\tilde{G}(k,z;z^{\prime })& =(2L/\beta )B(z^{\prime })^{-1}a^{1-D}(z^{\prime
})z^{\frac{D}{2}+\nu }z^{\prime }{}^{-\frac{D}{2}-\nu }e^{-\frac{(D-1)\eta
(z-z^{\prime })}{4L}} \\
& \left[ \theta (z-z^{\prime })\,_{1}F_{1}\left( \alpha +\nu ,2\nu +1;\frac{%
z^{\prime }\beta }{2L}\right) U\left( \alpha +\nu ,2\nu +1,\frac{z\beta }{2L}%
\right) \right.  \\
& \left. +\theta (z^{\prime }-z)_{1}F_{1}\left( \alpha +\nu ,2\nu +1;\frac{%
z\beta }{2L}\right) U\left( \alpha +\nu ,2\nu +1,\frac{z^{\prime }\beta }{2L}%
\right) \right] ,
\end{split}%
\end{equation}%
where
\begin{eqnarray}
B(z^{\prime }) &=&\frac{(\alpha +\nu )}{(2\nu +1)}\,\left( _{1}F_{1}\left(
\alpha +\nu +1,2\nu +2;\frac{z^{\prime }\beta }{2L}\right) U\left( \alpha
+\nu ,2\nu +1,\frac{z^{\prime }\beta }{2L}\right) \right.   \notag \\
&&+\left. (2\nu +1)\,_{1}F_{1}\left( \alpha +\nu ,2\nu +1;\frac{z^{\prime
}\beta }{2L}\right) U\left( \alpha +\nu +1,2\nu +2,\frac{z^{\prime }\beta }{%
2L}\right) \right) ,
\end{eqnarray}%
which comes from the denominator in (\ref{tildeG_ap}).

As long as $|\eta |~\ll 1,$ the two basis solutions (\ref{basis_01_1}), (\ref%
{basis_01_2}) have approximate forms
\begin{equation}
\phi _{1}(z)=z^{\frac{D}{2}}\exp \left( -\frac{(D-1)\eta }{4L}z\right)
K_{\nu }\left( (k^{2}+\xi ^{-2})^{\frac{1}{2}}z\right) ,
\end{equation}%
\begin{equation}
\phi _{2}(z)=z^{\frac{D}{2}}\exp \left( -\frac{(D-1)\eta }{4L}z\right)
I_{\nu }\left( (k^{2}+\xi ^{-2})^{\frac{1}{2}}z\right) ,
\end{equation}%
where $K_{\nu }\left( \cdot \right) $ and $I_{\nu }\left( \cdot \right) $
are the modified Bessel functions of the second and the first kind.
Therefore, the propagator is
\begin{equation}
\begin{split}
\tilde{G}(k,z;z^{\prime })=z^{D/2}(z^{\prime })^{1-\frac{D}{2}}e^{-\frac{%
(D-1)\eta (z-z^{\prime })}{4L}}& \left[ \theta (z-z^{\prime })K_{\nu }\left(
z\sqrt{k^{2}+\frac{1}{\xi ^{2}}}\right) I_{\nu }\left( z^{\prime }\sqrt{%
k^{2}+\frac{1}{\xi ^{2}}}\right) \right. \\
& \left. +\theta (z^{\prime }-z)K_{\nu }\left( z^{\prime }\sqrt{k^{2}+\frac{1%
}{\xi ^{2}}}\right) I_{\nu }\left( z\sqrt{k^{2}+\frac{1}{\xi ^{2}}}\right) %
\right] a^{1-D}(z^{\prime }).
\end{split}
\label{bulk_to_bulk_prop_ap}
\end{equation}

The boundary-to-bulk propagator $\tilde{K}(z,k)$ in momentum space could be
derived by taking limit $z^{\prime }\rightarrow 0$ in (\ref%
{bulk_to_bulk_prop_ap}) together with a normalization factor,
\begin{eqnarray}
\tilde{K}(z,k) &=&\lim_{z^{\prime }\rightarrow 0}2\nu (z^{\prime })^{\frac{D%
}{2}-1-\nu }a^{D-1}(z^{\prime })\tilde{G}(k,z;z^{\prime })  \notag \\
&=&\frac{2^{-\nu +1}}{\Gamma (\nu )}z^{D/2}e^{-\frac{(D-1)\eta z}{4L}}K_{\nu
}\left( z\sqrt{k^{2}+\frac{1}{\xi ^{2}}}\right) \left( \sqrt{k^{2}+\frac{1}{%
\xi ^{2}}}\right) ^{\nu }.
\end{eqnarray}%
Thus the boundary-to-bulk propagator $K(z,x;x^{\prime })$ in position space
could be gotten by Fourier transformation,
\begin{eqnarray}
&&%
\begin{split}
& K(z,x;x^{\prime })=\int \frac{\mathrm{d}^{D}\vec{k}}{(2\pi )^{D}}\tilde{K}%
(z,k)e^{i\vec{k}(\vec{x}-\vec{x}^{\prime })} \\
=& \frac{|x-x^{\prime }|^{1-\frac{D}{2}}}{(2\pi )^{D/2}}\int_{0}^{\infty }%
\mathrm{d}kk^{D/2}J_{\frac{D}{2}-1}(k|x-x^{\prime }|)\tilde{K}(z,k) \\
=& \frac{2^{-\nu +1}}{(2\pi )^{D/2}\Gamma (\nu )}\left( \frac{z}{\xi \sqrt{%
z^{2}+|x-x^{\prime }|^{2}}}\right) ^{\nu +\frac{D}{2}}e^{-\frac{(D-1)\eta z}{%
4L}}K_{\nu +\frac{D}{2}}\left( \xi ^{-1}\sqrt{z^{2}+|x-x^{\prime }|^{2}}%
\right) ,
\end{split}
\notag \\
&&
\end{eqnarray}%
which we present in Eq. (\ref{boundary_to_bulk_}).

\end{document}